\documentclass[prd, showpacs, amsmath,amssymb, amsmath]{revtex4}

\usepackage{dcolumn}
\usepackage{bm}

\begin{document}


\title{Quantum Equivalence of Massive Antisymmetric Tensor Field Models in Curved Space}

\author{I.L. Buchbinder}
\email{joseph@tspu.edu.ru}

\author{E.N.
Kirillova}
 \email{kirillovaen@tspu.edu.ru}
\affiliation{Department of Theoretical Physics\\ Tomsk State
Pedagogical University\\ Tomsk 634041, Russia.}

\author{N.G. Pletnev}
 \email{pletnev@math.nsc.ru}
 \affiliation{ Department of Theoretical Physics\\ Institute of Mathematics,
Novosibirsk, \\ 630090, Russia.}%

\date{\today}

\begin{abstract}
We study the effective actions in massive rank-2 and rank-3
antisymmetric tensor field models in curved space-time. These models
are classically equivalent to massive vector field and massive
scalar field with minimal coupling to gravity respectively. We prove
that the effective action for massive rank-2 antisymmetric tensor
field is exactly equal to that for massive vector field and the
effective action for massive rank-3 antisymmetric tensor field is
exactly equal to that for massive scalar field. The proof is based
on an identity for mass-dependent zeta-functions associated with
Laplacians acting on $p$-forms.

\end{abstract}



\maketitle

\section{Introduction}
Antisymmetric tensor fields or $p$-forms are the components of field
content of all superstring models and hence can survive in the
low-energy limit. Indeed, the attempts to relate Calabi-Yau
compactifications of ten-dimensional superstring theories to the
real world lead to low-energy supergravity theories with electric
and magnetic fluxes of various $p$-form fields. Such theories drew a
lot of attention due to observation that the resulting scalar
potentials in the low-energy effective supergravity theories might
lift the vacuum degeneracy \cite{L}, \cite{A}. In particular, the
magnetic charges should yield mass terms for $p$-forms \cite{AF}
and, in fact, we get the supersymmetric models containing the
massive antisymmetric tensor fields. Usually such fields belong to
tensor supermultiplets. The massive ${\cal N}=0,1,2$ tensor
multiplets \cite{LS}, \cite{K} possess some interesting properties.
However, the many aspects, especially the quantum ones, have not
been studied so far.

In this paper, as the first step in studying the quantum aspects of
massive supersymmetric theories we examine the structure of the
effective action in bosonic massive antisymmetric tensor field
models in four dimensional curved space-time. In four dimensions, a
massive rank-2 antisymmetric tensor field is classically equivalent
to a massive vector field. Similarly, a massive rank-3 antisymmetric
tensor field is equivalent to a massive scalar field minimally
coupled to gravity. In this paper we study the problem of quantum
equivalence of these classically equivalent theories.

Some years ago there was a large work on studying the massless
antisymmetric field models in curved space-time. In particular, in
four dimensions, a massless rank-2 antisymmetric field \cite{ogiev}
is classically equivalent to massless nonconformal scalar field and
massless rank-3 antisymmetric field has no physical degrees of
freedom\footnote{See discussion of the classical equivalence in
\cite{Her} and references herein.}. Quantization aspects were
discussed in \cite{schwarz1}, \cite{schwarz2}, \cite{Siegel},
\cite{sezgin}, \cite{kimura}, \cite{BK1}. The problem of quantum
equivalence of massless classically equivalent theories was
considered in \cite{schwarz2}, \cite{BK1}, \cite{Duff},
\cite{grisaru}, \cite{FT}. For massive antisymmetric field models,
the net of classical dualities in this case differs from the one for
massless fields. Therefore one can expect that studying the problem
of quantum equivalence for massive antisymmetric fields requires the
other methods in comparison with massless antisymmetric fields. Some
quantum aspects of massive antisymmetric field models have been
considered in \cite{bast} on the base of the worldline approach to
the effective action. It was pointed out in \cite{bast} that the
unregulated effective actions have a topological mismatch between
massive $p$-form and its dual massive $(D-p-1)$-form.

In this paper, we study a generic structure of the effective actions
for massive rank-2 and rank-3 antisymmetric fields in arbitrary
curved space-time and prove, using the zeta-function technique, that
they are exactly equal to those for massive vector field and massive
scalar field minimally coupled to gravity respectively. The proof is
essentially based on an identity for mass-dependent zeta-functions
associated with $p$-forms. The paper is organized as follows.
Section II is devoted to a brief description of massive
antisymmetric field models in four-dimensional curved space-time. In
Section III we discuss the definitions of the effective action for
massive antisymmetric field models. Section IV is devoted to proving
the  quantum equivalence of dual massive field models under
consideration. In Summary we formulate the results.

\section{Models of Massive Antisymmetric Tensor Fields}
We consider a model of massive antisymmetric second rank tensor
field $B_2=(B_{\mu\nu})$ in curved 4D space. The model is
described by the action
\begin{equation}
S[B_2]=\int
d^{4}x\sqrt{-g(x)}\left\{-\frac{1}{12}F^{\mu\nu\lambda}(B)F_{\mu\nu\lambda}(B)+
\frac{1}{4}\:m^{2}B^{\mu\nu}B_{\mu\nu}\right\}~, \label{action}
\end{equation}
where
\begin{eqnarray}
F_{\mu\nu\lambda}(B)=\nabla_{\mu}B_{\nu\lambda}+\nabla_{\nu}B_{\lambda\mu}+\nabla_{\lambda}B_{\mu\nu}~.
\label{F}
\end{eqnarray}
It is easy to see that the kinetic part of action (\ref{action})
is gauge invariant under the transformations
\begin{equation}
B_{\mu\nu}\rightarrow
B^{\xi}_{\mu\nu}=B_{\mu\nu}+\nabla_{\mu}\xi_{\nu} -
\nabla_{\nu}\xi_{\mu} \label{xi}
\end{equation}
with a vector gauge parameter $\xi_{\mu}$  defined up to a
transformation $\xi'_{\mu} = \xi_{\mu} + \nabla_{\mu}\xi$ with
scalar parameter $\xi$. This means that the gauge generators are
linearly dependent. The massive term in the action (\ref{action})
violates this symmetry.

For quantization of the theory and evaluation of the effective
action it is convenient \cite{BBS} to restore the gauge invariance
under (\ref{xi}) in massive theory (\ref{action}) with help of the
St\"uckelberg procedure. We introduce the vector field
$C_1=(C_{\mu})$ and consider the following action
\begin{equation}
S[B_2,C_1]=\int d^4 x
\sqrt{-g(x)}\{-\frac{1}{12}F^{\mu\nu\lambda}(B)F_{\mu\nu\lambda}(B)+
\frac{1}{4}\:m^{2}(B^{\mu\nu}+\frac{1}{m}F^{\mu\nu}(C))^2\}~,
\label{action1}
\end{equation}
where $F_{\mu\nu}(C) = \nabla_{\mu}C_{\nu}-\nabla_{\nu}C_{\mu}$.
The action (\ref{action1}) is invariant under the gauge
transformations (\ref{xi}) of the field 
$B_{\mu\nu} $ and under shift of the field $C_\mu$
\begin{equation}\label{gtBC}
 C_{\mu}\rightarrow C^{\xi}_{\mu}=C_{\mu}-m\xi_{\mu}
\end{equation}
and also under the gauge transformations of the St\"uckelberg
vector field:
\begin{equation}\label{gtCB}
C_{\mu}\rightarrow
C^{\Lambda}_{\mu}=C_{\mu}+\nabla_{\mu}\Lambda,\quad
B_{\mu\nu}\rightarrow B^{\Lambda}_{\mu\nu}=B_{\mu\nu}~,
\end{equation}
with a scalar gauge parameter $\Lambda.$ In four dimensions, a
massive rank-2 antisymmetric tensor field can be interpreted as a
massive pseudovector field. The physical component of $B_{\mu\nu}$
then corresponds to the longitudinal mode of a massive pseudovector
field, while the two physical components of $C_\mu$ correspond to
its transverse modes. The Lagrangian (\ref{action1}) can simply be
expressed as
\begin{equation}\label{tilde}
{\cal
L}=-\frac{1}{12}\widetilde{F}_{\mu\nu\lambda}\widetilde{F}^{\mu\nu\lambda}+\frac14
m^2\widetilde{B}_{\mu\nu}\widetilde{B}^{\mu\nu},\end{equation} where
$\widetilde{B}_{\mu\nu}\equiv B_{\mu\nu}+\frac{1}{m}F_{\mu\nu}(C)$
and $\widetilde{F}_{\mu\nu\lambda}$ is as in (\ref{F}). Since
$\widetilde{B}_{\mu\nu}$ is gauge invariant the gauge invariance of
${\cal L}$ is evident. Then it is obvious that ${\cal L}$ describes
on Abelian rank-2 antisymmetric tensor field with mass $m$. In four
dimensions the theory (\ref{action}) is classically equivalent to
the theory of massive vector field $A_{\mu}$  with action
\begin{equation}
S[A,C] =\int
d^{4}x\sqrt{-g(x)}\left\{-\frac{1}{4}F^{\mu\nu}(A)F_{\mu\nu}(A)+
\frac{1}{2}\:m^{2}(A^{\mu}-\frac{1}{m}\partial^\mu
C)(A_{\mu}-\frac{1}{m}\partial_\mu C)\right\}. \label{action2}
\end{equation}
Here $C$ is the St\"uckelberg scalar field. The equivalence is
resulted from the analysis of the equations of motion in both
theories. The duality relation looks like $mB_{\mu\nu}\sim
\epsilon_{\mu\nu\alpha\beta}F^{\alpha\beta}(A).$

Also we consider a model of massive totally antisymmetric third
rank tensor field $B_3=(B_{\mu\nu\rho})$ in curved space. Such a
model is described by the action
\begin{equation}
S[B_3]=\frac{1}{2}\int
d^{4}x\sqrt{-g(x)}\left\{-\frac{1}{4!}F^{\mu\nu\rho\sigma}(B)F_{\mu\nu\rho\sigma}(B)+
\frac{m^{2}}{3!}\:B^{\mu\nu\rho}B_{\mu\nu\rho}\right\},
\label{action3}
\end{equation}
where
\begin{eqnarray}
F_{\mu\nu\rho\sigma}(B)=\nabla_{\mu}B_{\nu\rho\sigma}-\nabla_{\nu}B_{\rho\sigma\mu}+
\nabla_{\rho}B_{\sigma\mu\nu}-\nabla_{\sigma}B_{\mu\nu\rho},
\label{F3}
\end{eqnarray}
$$B_{\nu\rho\sigma}=-B_{\rho\nu\sigma}=-B_{\nu\sigma\rho}=-B_{\sigma\rho\nu}.$$
The kinetic term of the action (\ref{action3}) is gauge invariant
under the transformations
\begin{equation}
B_{\mu\nu\rho} \rightarrow
B^{\xi}_{\mu\nu\rho}=B_{\mu\nu\rho}+\nabla_{\mu}\xi_{\nu\rho}+\nabla_{\nu}\xi_{\rho\mu}+\nabla_{\rho}\xi_{\mu\nu}
\label{xi3}
\end{equation}
with a tensor gauge parameter $\xi_{\mu\nu}=-\xi_{\nu\mu},$. This
parameter is defined up to a gauge transformation $\xi'_{\mu\nu} =
\xi_{\mu\nu}+ \nabla_{\mu}\xi_{\nu}-\nabla_{\nu}\xi_{\mu}$ with
vector a gauge parameter $\xi_{\mu}$. In its turn, the parameter
$\xi_{\mu}$ is defined up to gauge transformation $\xi'_{\mu} =
\xi_{\mu}+\nabla_{\mu}\xi$ with scalar gauge parameter $\xi$. This
means that the gauge generators are linearly dependent. As in the
previous case, we restore the gauge invariance under (\ref{xi3}) in
massive theory (\ref{action3}) with help of the St\"uckelberg
procedure. We introduce the second rank antisymmetric tensor field
$C_2=(C_{\mu\nu})$ and consider the following action
\begin{equation}
S[B_3,C_2]=\frac{1}{2}\int
d^{4}x\sqrt{-g(x)}\left\{-\frac{1}{4!}F^{\mu\nu\rho\sigma}(B)F_{\mu\nu\rho\sigma}(B)+
\frac{m^{2}}{3!}\:(B^{\mu\nu\rho}+\frac{1}{m}F^{\mu\nu\rho}(C))^2\right\}~,
\label{action4}
\end{equation}
where
\begin{equation}
F_{\mu\nu\rho}(C)=\nabla_{\mu}C_{\nu\rho}+\nabla_{\nu}C_{\rho\mu}+\nabla_{\rho}C_{\mu\nu}~.
\label{Fi3}
\end{equation}
The action (\ref{action4}) is invariant under the gauge
transformations of the fields $B^{\mu\nu\rho}$, $C_{\nu\rho}:$
\begin{equation}
B_{\mu\nu\rho} \rightarrow
B^{\xi}_{\mu\nu\rho}=B_{\mu\nu\rho}+\nabla_{\mu}\xi_{\nu\rho}+\nabla_{\nu}\xi_{\rho\mu}+\nabla_{\rho}\xi_{\mu\nu},\quad
C_{\mu\nu}\rightarrow C^{\xi}_{\mu\nu}=C_{\mu\nu}-m\xi_{\mu\nu}
\label{gtWX}
\end{equation}
and also under the St\"uckelberg gauge transformations
\begin{equation}\label{gtXW}
C_{\mu\nu}\rightarrow
C^{\Lambda}_{\mu\nu}=C_{\mu\nu}+\nabla_{\mu}\Lambda_{\nu}-\nabla_{\nu}\Lambda_{\mu},\quad
B_{\mu\nu\rho} \rightarrow
B^{\Lambda}_{\mu\nu\rho}=B_{\mu\nu\rho},
\end{equation}
because  $F^{\Lambda}_{\mu\nu\rho}=F_{\mu\nu\rho},$ where $\Lambda$
is a vector gauge parameter, defined up to a gauge transformation
$\Lambda'_{\mu} = \Lambda_{\mu}+ \nabla_{\mu}\Lambda$ with a scalar
parameter $\Lambda$. This means that the corresponding gauge
generators are linearly dependent. The equations of motion in the
theory (\ref{action4}) are equivalent to those in the theory
(\ref{action3}). In particular, the action (\ref{action4}) coincides
with the action (\ref{action3}) in the gauge $C_{\mu\nu}=0.$ One can
prove that in four dimensions the theory of a massive third rank
antisymmetric tensor field is classically equivalent to the theory
of a real massive scalar field $\phi$ minimally coupled to gravity.
The corresponding duality relation has the form
$mB_{\mu\nu\alpha\beta}\sim
\epsilon_{\mu\nu\alpha\beta}\partial^{\beta}\phi.$

\section{The Effective Action}

The models of massive antisymmetric fields in their initial
formulations (\ref{action}), (\ref{action3}) contain gauge
invariant kinetic terms and non-gauge invariant massive terms. In
this case the effective actions are given by the functional
determinants of the differential operators with degenerate
matrices at higher derivative terms but non-degenerate matrices at
mass terms. Then, the calculations of the effective actions
becomes very complicated and problematic\footnote{See discussion
of this point in \cite{BBS}.}. To avoid such a problem we
reformulated the models under consideration with help of the
St\"uckelberg procedure as the gauge theories. Therefore, to
construct the corresponding effective actions we can apply now the
quantization methods of gauge theories. It is especially worth
pointing out that the above models belong to a class of gauge
theories with linearly dependent generators. Their quantization is
very nontrivial and differs from quantization of the Yang-Mills
type theories where the gauge generators are independent. General
quantization procedure for theories with dependent generators in
the Lagrangian formalism is given by BV-method \cite{Batalin}.
However, quantization of simple theories with quadratic actions
and Abelian dependent gauge generators can be carried out by
successive multi-step applications of the Faddeev-Popov procedure
(see e.g. \cite{schwarz1}, \cite{schwarz2},
\cite{BK1})\footnote{The authors of \cite{Batalin} emphasized that
"To apply the methods of this paper to the simple linear theories
.... is like cracking nuts by a sledgehammer."}.

To quantize the theories under consideration we use a rather
simple procedure developed in \cite{BK1} (see also \cite{BK2}).
Omitting the calculations we formulate only the final results for
effective actions.

First, the effective action $\Gamma_{2}^{(m)}[g_{\mu\nu}]$ of the
massive second rank antisymmetric field model is given by the
relation
\begin{equation}
\Gamma_{2}^{(m)}[g_{\mu\nu}] =
\frac{i}{2}[\mbox{Tr}\ln(\Box_{2}+m^{2})-\mbox{Tr}\ln(\Box_{1}+m^{2})+\mbox{Tr}\ln(\Box_{0}+m^{2})]~.
\label{EE1}
\end{equation}

Second, the effective action $\Gamma_{1}^{(m)}[g_{\mu\nu}]$ of the
massive vector field $A_\mu$ is given by the relation
\begin{equation}
\Gamma_{1}^{(m)}[g_{\mu\nu}]=\frac{i}{2}[\mbox{Tr}\ln(\Box_{1}+m^{2})-\mbox{Tr}\ln(\Box_{0}+m^{2})]~.
\label{EE2}
\end{equation}

Third, the effective action $\Gamma_{3}^{(m)}[g_{\mu\nu}]$ of the
massive third rank antisymmetric tensor field is given by the
relation
\begin{equation}
\Gamma_{3}=\frac{i}{2}[\mbox{Tr}\ln(\Box_{3}+m^{2})-\mbox{Tr}\ln(\Box_{2}+m^{2})+\mbox{Tr}\ln(\Box_{1}+m^{2})-\mbox{Tr}\ln(\Box_{0}+m^{2})]~.
\label{EE3}
\end{equation}

Fourth, the effective action $\Gamma_{0}^{(m)}[g_{\mu\nu}]$ of the
massive scalar field $\phi$ with minimal coupling to gravity is
given by the relation
\begin{equation}
\Gamma_{0}^{(m)}[g_{\mu\nu}]=\frac{i}{2}\mbox{Tr}\ln(\Box_{0}+m^{2})~.
\label{EE4}
\end{equation}
Here the $\Box_{3}, \Box_{2}, \Box_{1}$ and $\Box_{0}$ are the
d'Alembertians acting on rank-$p$ antisymmetric tensor fields and
\begin{equation}
\mbox{Tr}(...) = \int d^{4}x\sqrt{-g(x)}\mbox{tr}(...)~,
\end{equation}
where $\mbox{tr}(...)$ is taken over tensor indices.

The relations (\ref{EE1}), (\ref{EE2}), (\ref{EE3}), (\ref{EE4})
can also be understood in terms of dimensional reduction
\cite{Scherk}. Let us consider massless antisymmetric tensor
fields $B_{MN}$ and $B_{MNK}$ in five-dimensional space of the
topology $R_4 \times S^1$ where is $R_4$ is the four-dimensional
Riemannian space. In D5, the massless field $B_{MN}$ has three
physical degrees of freedom and equivalent to a massless vector,
the massless field $B_{MNK}$ has one physical degrees of freedom
and equivalent to a massless scalar. General form of effective
actions of massless antisymmetric tensor field models has been
obtained in \cite{Siegel}, \cite{BK1}, \cite{Duff},  \cite{bast}.
The results for the corresponding five-dimensional effective
actions are
$$\Gamma^{(D5)}_2=\frac{i}{2}[\mbox{Tr}\ln(\Box_{2})-2\mbox{Tr}\ln(\Box_{1})+3\mbox{Tr}\ln(\Box_{0})]~,$$
$$\Gamma^{(D5)}_3=\frac{i}{2}[\mbox{Tr}\ln(\Box_{3})-2\mbox{Tr}\ln(\Box_{2})+3\mbox{Tr}\ln(\Box_{1})-4\mbox{Tr}\ln(\Box_{0})]~.$$
Here all $\mbox{Tr}$-operations are defined in D5. Then, following
\cite{Scherk} we identify $B_{MN}$ as $B_{\mu\nu}, C_{\mu}$ and
$B_{MNK}$ as $B_{\mu\nu\rho}, C_{\mu\nu}$ and specify the $x_5$
dependence by setting $e^{imx_5}$. After that, $\Gamma^{(D5)}_2$
and $\Gamma^{(D5)}$ exactly reproduce (\ref{EE1}) and (\ref{EE3})
respectively \footnote{Reduction to D4 yields to
$\mbox{Tr}\ln(\Box_{3})=\mbox{Tr}\ln(\Box_{3}+m^2)+
\mbox{Tr}\ln(\Box_{2}+m^2)$,
$\mbox{Tr}\ln(\Box_{2})=\mbox{Tr}\ln(\Box_{2}+m^2)+
\mbox{Tr}\ln(\Box_{1}+m^2)$ and $\mbox{Tr}\ln(\Box_{1})=
\mbox{Tr}\ln(\Box_{1}+m^2)+\mbox{Tr}\ln(\Box_{0}+m^2)$. Here left
hand sides correspond to D5 and right hand sides to D4.}.

\section{Quantum Equivalence}
To study the quantum equivalence of the classically equivalent
theories one considers the differences of the effective actions
(\ref{EE1}) and (\ref{EE2}) and also (\ref{EE3}) and (\ref{EE4}).
Denoting
\begin{equation}
\Delta\Gamma^{(1)} = \Gamma_{2}^{(m)}[g_{\mu\nu}] -
\Gamma_{1}^{(m)}[g_{\mu\nu}] \label{delta1}
\end{equation}
and
\begin{equation}
\Delta\Gamma^{(2)} = \Gamma_{3}^{(m)}[g_{\mu\nu}] -
\Gamma_{0}^{(m)}[g_{\mu\nu}] \label{delta2}
\end{equation}
and using the relations (\ref{EE1}), (\ref{EE2}), (\ref{EE3}), and
(\ref{EE4}) ones get
\begin{equation}
\Delta\Gamma^{(1)}=\frac{i}{2}\left[\mbox{Tr}\ln(\Box_{2}+m^{2})-2\mbox{Tr}\ln(\Box_{1}+m^{2})+
2\mbox{Tr}\ln(\Box_{0}+m^{2})\right]~, \label{differ1}
\end{equation}
\begin{equation}
\Delta\Gamma^{(2)}=\frac{i}{2}[\mbox{Tr}\ln(\Box_{3}+m^{2})-\mbox{Tr}\ln(\Box_{2}+m^{2})+
\mbox{Tr}\ln(\Box_{1}+m^{2})-2\mbox{Tr}\ln(\Box_{0}+m^{2})]~.
\label{differ2}
\end{equation}

It is easy to see, taking into account the numbers of independent
components of the antisymmetric third and second rank tensors, that
in flat space the relation (\ref{differ1}) is exactly zero.
Similarly, the relation (\ref{differ2}) is also zero. Now we will
study the structure of the relations (\ref{differ1}) and
(\ref{differ2}) in curved space and prove, using the definition of
functional determinants in terms of generalized $\zeta$-function,
that they are still zero. This means that the above classically
equivalent theories are quantum equivalent. Before presenting a
proof, we point out that the relations (\ref{differ1}),
(\ref{differ2}) contain mass $m$ and, therefore, a proof of quantum
equivalence should include some new issues in comparison with the
corresponding massless cases.

It is convenient to consider the fields $\phi, A_\mu, B_2$ and $B_3$
as the corresponding $p$-forms $(p=0,1,2,3)$. In Euclidean
formulation,the operators $\Box_p$ become the corresponding
Laplacians, acting on $p$-forms (see the details in \cite{N},
\cite{R}). We define the effective actions in terms of generalized
zeta-functions $\zeta_{p}(s,m)$ \footnote{See the various
applications of zeta-function techniques in \cite{eliza}.}
associated with the operators $-\Box_{p} + m^2$
\begin{equation}
\zeta_{p}(s,m) =\sum_{\lambda_i\neq 0}\lambda_i^{-s}=
\frac{1}{\Gamma(s)}\int_0^\infty dtt^{s-1}
e^{-tm^{2}}\mbox{Tr}(e^{t\Box_{p}}-{\cal P}_{p})~, \label{zeta}
\end{equation}
where ${\cal P}_{p}$ is the projector onto the space of the zero
modes of the operator $\Box_{p}$ \cite{Ray}, \cite{R}. According to
this definition the zeta-function is analytic at $s=0$ together with
its derivative. In these terms, the effective action associated with
the operator $(-\Box_{p} + m^2)$ is given by
\begin{equation}
\mbox{ln}\mbox{Det}(-\Box_{p} + m^2) = -(\zeta_{p}'(0,m) +
\mbox{ln}({\mu}^2)\zeta_{p}(0,m))~. \label{zeta1}
\end{equation}
Here $\mu$ is an arbitrary mass scale parameter. By definition, the
expression (\ref{zeta1}) is finite  for any $p$. Using the relation
(\ref{zeta1}) we rewrite  (\ref{differ1}) and (\ref{differ2}) in the
form
\begin{equation}
\Delta\Gamma^{(1)}=
(\zeta'_{2}(0,m)-2\zeta'_{1}(0,m)+2\zeta'_{0}(0,m)) +
\mbox{ln}({\mu}^2)(\zeta_{2}(0,m)-2\zeta_{1}(0,m)+2\zeta_{0}(0,m))
\label{delta3}
\end{equation}
and
\begin{equation}
\Delta\Gamma^{(2)}= (\zeta'_{3}(0,m)-\zeta'_{2}(0,m)+
\zeta'_{1}(0,m)-2\zeta'_{0}(0,m)) \label{delta4}
\end{equation}
$$+
\mbox{ln}({\mu}^2)(\zeta_{3}(0,m)-\zeta_{2}(0,m)+\zeta_{1}(0,m)-2\zeta_{0}(0,m))~.$$
Taking into account the Hodge duality between $p$-form and
$(4-p)$-form and the corresponding properties of the operators
$\Box_{p}$
 \cite{N}, where $p=0,1,2,3,4$, ones get
$\zeta_{p}(s,m) = \zeta_{(4-p)}(s,m)$ for any $m$\footnote{See
proof of this identity for $m=0$ in \cite{R}.}. Then it is
evident, that
\begin{equation}
\Delta\Gamma^{(2)} = -\Delta\Gamma^{(1)}~. \label{1=-2}
\end{equation}
Hence, it is sufficient to study only $\Delta\Gamma^{(1)}$
(\ref{delta3}).

Let us apply the evident identity
\begin{equation}
\sum_{p=0}^{4} (-1)^{p}p\;\zeta_{p}(s,m)=
2(\zeta_{2}(s,m)-2\zeta_{1}(s,m)+2\zeta_{0}(s,m))
\label{identity1}
\end{equation}
to the expression (\ref{delta3}). Then one obtains
\begin{equation}
\Delta\Gamma^{(1)} = \frac{1}{2}[\sum_{p=0}^{4}
(-1)^{p}p\;\zeta'_{p}(0,m)+\mbox{ln}({\mu}^2)\sum_{p=0}^{4}
(-1)^{p}p\;\zeta_{p}(0,m)]~. \label{delta5}
\end{equation}
Expansion of the zeta-function at non-zero mass (\ref{zeta}) in
power series in mass allows us to get the mass-dependent
zeta-function as a power series in massless zeta-functions in the
form
\begin{equation}
\zeta_{p}(s,m) = \sum_{n=0}^\infty
\frac{(-m^2)^{n}\Gamma(n+s)}{n!\Gamma(s)}\zeta_{p}(s+n,0)~.
 \label{expansion}
\end{equation}
It allows us to write
\begin{equation}
\sum_{p=0}^{4} (-1)^{p}p\;\zeta_{p}(s,m)=\sum_{n=0}^\infty
\frac{(-m^2)^{n}\Gamma(n+s)}{n!\Gamma(s)}\sum_{p=0}^4
(-1)^{p}p\;\zeta_{p}(s+n,0)~. \label{identity2}
\end{equation}
However, the zeta-functions at zero mass $\zeta_{p}(s+n) =
\zeta_{p}(s+n,0)$ satisfy for any $n$ the identity \cite{Ray} (see
also \cite{R} for a review)
\begin{equation}
\sum_{p=0}^4 (-1)^{p}p\;\zeta_{p}(s)=0~. \label{identity3}
\end{equation}
The equations (\ref{identity2}) and (\ref{identity3}) mean that
the analogous identity is also valid for mass-dependent
zeta-functions
\begin{equation}
\sum_{p=0}^4 (-1)^{p}p\;\zeta_{p}(s,m)=0~. \label{identity4}
\end{equation}
Using this relation in (\ref{delta5}) ones get
\begin{equation}
\Delta\Gamma^{(1)} = 0~. \label{fin}
\end{equation}
The last relation means that the effective action of a massive
second rank antisymmetric tensor field coincides with the
effective action of a massive vector field. That is these theories
are quantum equivalent. Taking into account (\ref{1=-2}) one
concludes that the effective action of a massive third rank
antisymmetric tensor field coincides with the effective action of
a massive scalar field. That is these theories are also quantum
equivalent.

Usually, the (one-loop) effective actions, associated with
differential operators, are defined in quantum field theory with
help of the Schwinger-De Witt representation. It is equivalent to
using another zeta-function, which includes the zero modes. Such
zeta-function is defined by (\ref{zeta}) where now the term with the
projector ${\cal P}_{p}$ is omitted (see e.g. \cite{H},
\cite{eliza}). Then the relation (\ref{1=-2}) still holds. However,
the identity (\ref{identity3}) is not valid now and, hence, the
identity (\ref{identity4}) is also not valid. Therefore, the final
relation (\ref{fin}) would be violated. First of all, we point out
that the relation (\ref{1=-2}) will still be valid if we define the
effective actions on the base of the zeta-function including the
zero modes. Relation (\ref{differ1}) can be identically rewritten in
terms of the zeta-functions including the zero modes. The difference
of the two zeta-functions is given by $(m^2)^{s}\mbox{Tr}{\cal
P}_{p}$, which is exactly calculated in terms of De Witt
coefficients at the coincident limit, associated with the operators
$\Box_{p}$. As a result one obtains
\begin{equation}
\Delta{\tilde \Gamma^{(1)}} =
\mbox{ln}(\frac{{\mu}^2}{m^2})\frac{1}{16\pi^{2}}\int
d^{4}x\sqrt{-g(x)} [b_{2}-m^2b_{1}+\frac{m^4}{2}b_{0}]~,
\label{fin1}
\end{equation}
where ${\tilde \Gamma}$ means the effective action defined in terms
of zeta-function, including the zero modes. Here
\begin{equation}\label{b}
b_{n} = a_{n}^{(2)}-2a_{n}^{(1)}+2a_{n}^{(0)}, \quad   n=0,1,2~.
\end{equation}
The De Witt coefficients at coincident limit $a^{(p)}_{n}, n=0,1,2;
p=0,1,2$ are known in literature (see e.g. \cite{DW}, \cite{bast}
and reference therein). Using the results of these calculations ones
get: ${\bf (i).}$ The coefficients $b_{0}=0$ owing to the elementary
balance of degrees of freedom for antisymmetric tensor fields. ${\bf
(ii).}$ The coefficients $b_{1}=0.$ This means that the difference
of the effective action for classically equivalent massive theories
is mass independent. ${\bf (iii).}$ Taking into account the explicit
expressions for the coefficients $a_{2}^{(p)}, p=0,1,2$, ones obtain
\begin{equation}
b_{2} = \frac{1}{2}[R_{\mu\nu\lambda\rho}^{2} - 4R_{\mu\nu}^{2} +
R^{2}].
\end{equation}
\label{topol1} As a result we get
\begin{equation}\label{topol2}
\Delta{\tilde \Gamma^{(1)}}=\mbox{ln}(\frac{{\mu}^2}{m^2})\chi,
\end{equation}
where $\chi$ is the Gauss-Bonnet topological invariant.  The
effective actions, defined in terms of the zeta-function
(\ref{zeta}) and in terms of the zeta-function with the zero modes
are equal to each other  up to the topological invariant. The
corresponding energy-momentum tensors coincide. If we define a
quantum equivalence of the theories as equality of their currents
(energy-momentum tensors in the given case), the definitions of the
effective actions both in terms the zeta-function (\ref{zeta}) and
in terms of the zeta-function with zero modes, lead to the same
conclusion on the quantum equivalence.

\section{Summary}

We have studied the structure of the effective actions in massive
second rank and third rank antisymmetric tensor field models in
four-dimensional curved space-time. The effective actions are
defined in terms of the generalized zeta-function associated with
the corresponding d'Alembertians (\ref{zeta}). We have proven that
the effective action for a massive second rank antisymmetric tensor
field is exactly equal to the effective action for a massive vector
field. Similarly, we have shown that the effective action for a
massive third rank antisymmetric tensor field is exactly equal to
the effective action for a massive scalar field minimally coupled to
gravity. The proof is essentially based on the identity
(\ref{identity4}) for mass-dependent zeta-functions (\ref{zeta}).
Our general statement is analogous to the one of \cite{bast}
although our method is quite different.

We would like to point out that the zeta-function $\zeta_{p}(s,m)$
(\ref{zeta}) in the basic identity (\ref{identity4}), does not
contain the zero modes of the operators $\Box_{p}$ unlike another
zeta-function, which is used often for definitions of the effective
actions. Two these definitions of the zeta-function differ by the
quantity $(m^2)^{s}\mbox{Tr}{\cal P}_{p}$. We have shown that if the
effective action is defined in terms of the zeta-function with the
zero modes the difference between the effective actions of
classically equivalent theories under consideration is the
Gauss-Bonnet topological invariant. This means that the effective
energy momentum tensors for these two theories coincide. Treating
the quantum equivalence of two theories as the equality of their
effective energy-momentum tensors, we conclude that the given
classically equivalent theories are quantum equivalent both if the
effective action is defined in terms of the zeta-function
(\ref{zeta}) and if it is defined in terms of the zeta-function with
the zero modes.

\section*{Acknowledgments}

We are grateful to F. Bastianelli for bringing to results of the
paper \cite{bast} to our attention and to E. Elizalde for useful
comments. The research was partially supported by LRSS grant,
project No. 2553.2008.2. Work of I.L.B and N.G.P was partially
sponsored by RFBR grant, project No. 06-02-16346, No.
08-02-00334-a and INTAS grant, project No. 05-1000008-7928. Work
of I.L.B was also partially supported by joint RFBR-Ukraine grant,
project No. 08-02-90490.

\end{document}